\documentclass[10pt, conference, compsocconf]{IEEEtran}
\usepackage{cite}
\usepackage{amsmath,amssymb,amsfonts}
\usepackage{algorithmic}
\usepackage{graphicx}
\usepackage{textcomp}
\usepackage{xcolor}
\usepackage[hyphens]{url}
\usepackage[ruled,lined]{algorithm2e}
\usepackage{mathtools}
\usepackage{subfig}
\usepackage{pifont}
\usepackage{multirow}
\usepackage{tablefootnote}
\usepackage{listings}
\usepackage{framed}
\usepackage{capt-of}
\usepackage{newfloat}
\usepackage[scaled]{beramono}
\usepackage{color}
\usepackage{flushend}

\definecolor{keywordcolor}{RGB}{157,0,129}
\definecolor{commentcolor}{RGB}{157,0,129}
\definecolor{numbercolor}{RGB}{120,120,120}
\DeclareFloatingEnvironment[fileext=frm,placement={!ht},name=Listing]{listing}

\newcommand{\myline}{\vspace{-0.7\baselineskip}\hrulefill\vspace{-0.1\baselineskip}}

\ifCLASSINFOpdf
\else
\fi
\begin{document}
\title{Abusing Cache Line Dirty States to Leak Information in Commercial Processors}


\author{\IEEEauthorblockN{Yujie Cui,
		Chun Yang, Xu Cheng}
\IEEEauthorblockA{Department of Computer Science \& Technology,  \\
	Engineering Research Center of Microprocessor \& System, \\
 Peking University, Beijing, China\\
Email: YujieCui@pku.edu.cn,
yangchun@mprc.pku.edu.cn,
chengxu@mprc.pku.edu.cn}
%\and
%\IEEEauthorblockN{Authors Name/s per 2nd Affiliation (Author)}
%\IEEEauthorblockA{line 1 (of Affiliation): dept. name of organization\\
%line 2: name of organization, acronyms acceptable\\
%line 3: City, Country\\
%line 4: Email: name@xyz.com}
}


%Telephone: (800) 555--1212, Fax: (888) 555--1212}


\maketitle

\begin{abstract}
Caches have been used to construct various types of covert and side channels to leak information. Most existing cache channels exploit the timing difference between cache hits and cache misses. However, we introduce a new and broader classification of cache covert channel attacks: Hit+Miss, Hit+Hit, and Miss+Miss. We highlight that cache misses (or cache hits) for cache lines in different states may have more significant time differences, and these can be used as timing channels. Based on this classification, we propose a new stable and stealthy Miss+Miss cache channel. Write-back caches are widely deployed in modern processors. This paper presents in detail a way in which replacement latency differences can be used to construct timing-based channels (called WB channels) to leak information in a write-back cache. Any modification to a cache line by a sender will set it to the dirty state, and the receiver can observe this through measuring the latency of replacing this cache set. We also demonstrate how senders could exploit a different number of dirty cache lines in a cache set to improve transmission bandwidth with symbols encoding multiple bits. The peak transmission bandwidths of the WB channels in commercial systems can vary between 1300 and 4400~kbps per cache set in a hyper-threaded setting without shared memory between the sender and the receiver. In contrast to most existing cache channels, which always target specific memory addresses, the new WB channels focus on the cache set and cache line states, making it difficult for the channel to be disturbed by other processes on the core, and they can still work in a cache using a random replacement policy. We also analyzed the stealthiness of WB channels from the perspective of the number of cache loads and cache miss rates. We discuss and evaluate possible defenses. The paper finishes by discussing various forms of side-channel attack.

\end{abstract}


% \fi
\IEEEpeerreviewmaketitle

\section{Introduction}

In recent years, computer security has attracted significant attention, and security and trustworthiness have become essential factors in the design of modern processor hardware. Leakage channels are classified according to the threat model: side channels refer to the accidental leakage of sensitive data by a trusted party, while covert channels use internal trojan processes to stealthily deliver secret information to a spy even when the underlying system security policy explicitly prohibits any such activity \cite{DepartmentofDefense1985}. Many side and covert channels have been used to break various encryption algorithms and leak private keys \cite{Gullasch2011,Percival2005,Bonneau2006,Osvik2006a}. Among these covert and side channels, time-based channels are the most frequently used and the most difficult to detect and defend against.  
Caches are the regularly used resources for attackers to build
time-based channels, because different cache behaviors have
rich time characteristics that are relatively easy to collect.

%Caches are the most regularly used resources for attackers to build time-based channels, because different cache behaviors have rich time characteristics that are relatively easy to collect.

To date, there has been a large amount of research using caches to construct timing-based channels and leak private data \cite{Osvik2006a,Yarom2014,Xiong2020c,Briongos2019,Liu2015}. The majority of these exploit the timing difference between cache hits and cache misses. However, we believe that cache misses (or cache hits) for cache lines in different \emph{states also have time differences that can be used as covert timing channels}. Based on this view, we focus on other time characteristics of the cache (e.g., cache misses). 
%covert timing channels

\begin{figure}[t]
	\centering
	\includegraphics[width=\linewidth]{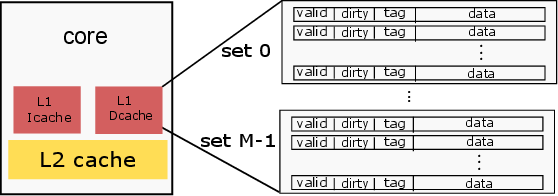}
	\caption{Cache organization and the phases of the new WB timing-based side and covert channels.}
	\label{CacheSet}
\end{figure}

In processor caches, the replacement algorithm determines where a loaded cache line should be placed when there is a cache miss. As shown in Figure~\ref{CacheSet}, a dirty bit is maintained for each cache line in a cache set in a write-back cache. This is used to determine whether the replaced cache line (called the \textit{victim way}) needs to be written back to the next-level cache or the main memory, which affects the completion time of the replacement. Based on this observation, we investigate and present a new type of cache channel that is exposed by the difference in replacement latency according to cache line states in a cache set without shared memory between the sender and receiver, referred to as the \textit{write-back} (WB) channel.

Previous time-based cache channels have always targeted specific memory addresses\cite{Yarom2014,Gruss2016a,Xiong2020c}. These generally either require shared memory between the sender and receiver or are easily disturbed by cache lines loaded by other parts of the program or other processes on the core. However, the WB channel focuses on the cache set and cache-line status, making the channel more robust and stealthier. Moreover, all cache lines in a cache set can be used equally, which allows us to exploit multi-bit encoding to increase the transmission rate. Furthermore, many secure caches \cite{Brickell2006,Fang2021} are designed for attacks exploiting the timing difference between cache hits and cache misses and are ineffective against the new WB channel.

In this paper, we present an investigation of a new type of hardware vulnerability to time-based channels exposed by the difference in replacement latency according to the state of a cache line in a cache set. Based on this vulnerability, we implemented the WB channel and demonstrated the attack on a commercial system. To accurately observe the time for the cache lines to be replaced, dedicated data structures and a pointer-chasing algorithm were used in the receiver's program to allow fine-grained memory-access latency measurements. Two algorithms were designed to establish WB time-based channels that transmit one or more bits at a time. To summarize, this paper makes the following contributions:

\begin{itemize}
	\setlength{\itemsep}{0pt}
	\setlength{\parsep}{0pt}
	\setlength{\parskip}{0pt}
	\item We propose a new classification of cache covert-channel attacks: Hit+Miss, Hit+Hit, and Miss+Miss.
	\item We present a way in which the dirty bits in cache lines can be used as timing-based cache channels for leaking information without shared memory between the sender and receiver.
	\item We analyze and evaluate the transmission rates and bit error rates of WB covert channels in detail. We also demonstrate how senders could exploit a different number of dirty cache lines in a target set to improve transmission bandwidth with symbols encoding multiple bits. We have effectively increased the achievable peak bandwidth from 1300~kbps with binary-encoded symbols to approximately 4400~kbps with multi-bit encoding.
	\item We compare the WB channel with existing cache channels from the perspectives of stability and stealthiness. We also demonstrate that the WB channel can still work in a cache using a random replacement policy.
	\item We propose and evaluate possible defenses against this type of attack and discuss how the new WB channels can be used to construct possible side-channel attacks.
\end{itemize}

\section{Background and Terminology}

\subsection{Classification of Cache Covert Channels}\label{chapter:classification}
Cache channel attacks have been conventionally classified as either contention-based attacks or reuse-based attacks \cite{Liu2015a} based on how the attacker infers the memory address. In contention-based attacks, the attacker contends with the victim process for the same cache set, and this contention causes the cache line of one to be evicted by that of the other. In reuse-based attacks, the attacker leverages the fact that previously accessed data will be cached, so reusing the same data in later memory accesses will result in cache hits. Therefore, reuse-based attacks generally rely on shared memory between the attacker and the victim. Page classified cache channel attacks as being either access driven or timing driven based on what can be measured by the attacker \cite{Page2002}. \textit{However, these classifications are essentially for attacks exploiting the timing difference caused by whether the data is cached.} We highlight that cache misses (or cache hits) for cache lines in different states also have time differences that can be used to construct covert channels. Therefore, we introduce new classifications: Hit+Miss, Hit+Hit, and Miss+Miss. These new classifications provide a broader understanding of cache attacks and prompt us to pay more attention to the details of the cache behavior to discover the sources of hardware vulnerabilities. Table~\ref{table:classification} summarizes the classifications of known cache covert-channel attacks. 
We have extended the classifications of cache channels proposed by Liu et~al. \cite{Ssociativity1997}, which are no longer limited to Hit+Miss attacks caused by data contention or reuse. 
%The operation sequences of cache hits and misses can be described as follows.

\begin{table}[t]
	\caption{Classification of cache covert channels.}
	\label{table:classification}
	\centering
	\resizebox{0.45\textwidth}{!}{
		\begin{tabular}{|c|c|c|c|}
			\hline
			& \textbf{\begin{tabular}[c]{@{}c@{}}Hit+Miss \\ Attacks\end{tabular}} & \textbf{\begin{tabular}[c]{@{}c@{}}Hit+Hit \\ Attacks\end{tabular}} & \textbf{\begin{tabular}[c]{@{}c@{}}Miss+Miss \\ Attacks\end{tabular}} \\ \hline
			\textbf{\begin{tabular}[c]{@{}c@{}}Contention-based\\ attacks\end{tabular}} & \begin{tabular}[c]{@{}c@{}}Prime+Probe\\ LRU\tablefootnote{\label{LRU}Xiong et~al. \cite{Xiong2020c} present least-recently used (LRU) timing-based channels both when the sender and the receiver do and do not have shared memory.}\\ Evict+Time\end{tabular} & \begin{tabular}[c]{@{}c@{}}CacheBleed \\ TLBleed \end{tabular} & \begin{tabular}[c]{@{}c@{}}Cache-replacement \\ latency (WB)\end{tabular} \\ \hline
			\textbf{\begin{tabular}[c]{@{}c@{}}Reuse-based \\ attacks\end{tabular}} & \begin{tabular}[c]{@{}c@{}}Flush+Reload\\ Flush+Flush\\ Evict+Reload\\ LRU\textsuperscript{\ref{LRU}}\end{tabular} & & \begin{tabular}[c]{@{}c@{}}Cache coherent \\ protocol states\end{tabular} \\ \hline
		\end{tabular}
	}
\end{table}

\begin{figure*}[t]
	\centering
	\includegraphics[width=\linewidth]{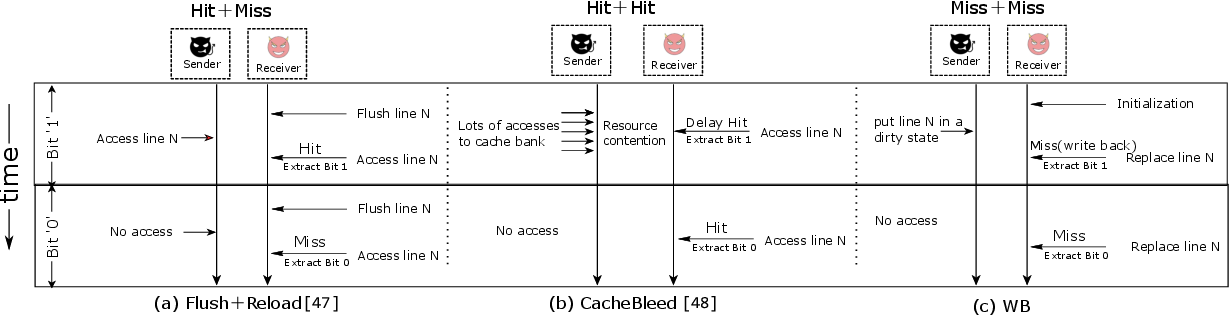}
	\caption{Examples of cache covert channel classifications.}
	\label{classification}
\end{figure*}

In Hit+Miss attacks, the attacker affects whether a specific cache line is available in the cache to leak information. The attacker can put a cache line into the cache by issuing a \textit{load} (or \textit{store}) instruction\cite{Yarom2014}, prefetching \cite{Gruss2016}, or speculatively fetching \cite{Evtyushkin2018,ChowdhuryyMdHafizulIslamandLiuHangandYao2020,Evtyushkin2016b}, and invalidating a cache line through cache contention \cite{Liu2015,Osvik2006a} or the \textit{clflush} instruction \cite{Gruss2016a,Yarom2014}. For example, Yarom and Falkner \cite{Yarom2014} exploited the latency difference between cache hits and DRAM accesses to establish a high-resolution covert channel on the last-level cache (LLC). Moreover, a profound understanding of the cache replacement policy \cite{Briongos2019,Xiong2020c} allows the attacker to more precisely control whether a specific cache line is available in the cache.

\begin{itemize}
	\setlength{\itemsep}{3pt}
	\setlength{\parsep}{0pt}
	\setlength{\parskip}{0pt}
	\item Cache hits:
	\ding{172} Translate virtual addresses (translation lookaside buffer, page table) $\to$ \ding{173} Access the cache (cache bank contention) $\to$ \ding{174} Send data to CPU (bus contention).
	\item Cache misses:
	\ding{172} Issue load requests (miss-status holding registers) $\to$ \ding{173} Select the victim cache line $\to$ \ding{174} Fetch the new cache line from the backing store (data location, cache coherency states) $\to$ \ding{175} Replace the victim cache line (victim cache line states) $\to$ \ding{176} Send data to CPU (bus contention).
\end{itemize}

As shown in the operation sequences above, cache hits and cache misses can also be regarded as micro-operations. Affecting the time for micro-operations to complete can create time variations in cache hits (or cache misses). Note that many micro-operations may be conducted in parallel. The parentheses in the above sequences point out the sources that may affect the completion time of these micro-operations. For example, Yarom et~al. exploited the creation of contention on a cache bank\footnote{To facilitate concurrent access to the cache, the Intel L1 cache comprises multiple banks. Concurrent accesses to different cache banks can always be served. However, each bank can only handle one access request at a time.}, measuring the timing variations due to the conflict to establish a brand new cache channel (Hit+Hit) \cite{Yarom2017}. Yao et~al. revealed how to manipulate the coherence states on shared cache blocks to affect the time taken to fetch a new cache line and construct covert timing channels (Miss+Miss) \cite{Yao2018}. Based on the impact of this classification, this paper exploits using cache line states to affect the cache replacement latency (i.e., Miss+Miss) to establish a new cache channel. Brief descriptions of three covert channels (Flush+Reload \cite{Yarom2014}, CacheBleed \cite{Yarom2017}, and WB) under this classification are shown in Figure~\ref{classification}.

Hit+Miss attacks always require either shared memory between the sender and receiver or a profound understanding of the cache replacement policy, and they are easily disturbed by cache lines loaded by other parts of the program or other processes on the core. Hit+Hit attacks such as CacheBleed always require the sender and receiver to be two concurrent hyper-threads, making them challenging to deploy. Furthermore, these attacks always target specific memory addresses. However, this paper focuses on cache set and cache line states and exploits the replacement latency difference to establish stable and stealthy cache channels.

\subsection{Cache Write Policy}
In modern architecture, caches are essential components to hide memory access latency, as the clock rates of processors and the latency of memory have diverged dramatically in the last three decades \cite{Xiong}. When the CPU wants to write data to a cache, it will first check the tag to see if the address is a hit. The data in the cache will be modified if a cache hit appears. However, it must also write this data to the backing store at some point. The timing of this write is determined by what is known as the \textit{write policy}, which is typically divided into two basic approaches: write-through and write-back.

\textbf{Write-through.} When the CPU performs store operations, data is updated to the cache and main memory synchronously. The advantages of a write-through cache are that it is easy to implement and the data in the cache and main memory are consistent. However, write operations will experience high latency as the CPU has to write to the slower main memory. 
Therefore, the write-through policy has not been widely adopted.

\textbf{Write-back.} The data is updated only in the cache and updated into the next level cache or main memory when the cache line is replaced. As shown in Figure~\ref{CacheSet}, each block in the cache set needs a \textit{dirty bit} to indicate whether the cache line in the set was modified (dirty) or not modified (clean). When a cache miss appears, the replacement policy selects a cache line from the cache set to replace. A clean cache line can be directly replaced to reduce write operations. However, replacing a dirty cache line requires the backing store to be updated first. A write-back policy significantly reduces the write waiting time and the number of write operations, which improves the efficiency of the CPU. Therefore, the write-back policy is generally deployed in current processors.

Since no data is returned to the requester during the write operations, a decision needs to be made on write misses. Write misses can be divided into two policies according to whether the data is loaded into the cache when they occur: \textit{write allocate} or \textit{no-write allocate}. Either write-miss policy can be used with write-through or write-back. However, write-back caches typically use the write allocate policy, hoping that subsequent write operations to this block will only need to modify the cache. Write-through caches generally use the no-write allocate policy because subsequent write operations to this block still need to go to the memory.

\begin{algorithm} [t]
	\caption{Sender Operations in WB Channel}
	lines 0--$N$: cache lines mapping to the target set
	
	$m$: 1-bit information to transfer
	
	$M[k]$: $k$-bit message
	
	$d$: a parameter of the sender
	
	\myline
	
	\textbf{Encoding binary symbol:}
	
	\myline
	
	// Encoding Phase \newline
	\eIf{$m==1$}{
		Modify line $N$; // Put line $N$ in the dirty state

	}{	
		no access;
	}

	sleep; // Allow the receiver code to decode
	
	\myline
	
	\textbf{Encoding multi-bit symbol:}
	
	\myline
	
	// Encoding Phase \newline
	$d = f(M[0] - M[k-1]$);
	
	\For {$j=0$; $j<d$; $j$++}{
		Modify line $j$;	// Put line $j$ in the dirty state
	}
	
	sleep; 
	// Allow the receiver code to decode
\end{algorithm}

\section{Threat Model and Assumptions}
In this paper, we demonstrate how to use the replacement latency difference to construct our covert channels. When the victim has secret-dependent data access, covert channels can be extended to side channels \cite{Bonneau2006,Briongos2019,Yarom2014,Liu2015,Yan2019}.

In our WB time-based channels, we assume $N$-way write-back caches using the write allocate policy, which means that the cache line will only be updated to the next-level cache or main memory when it is replaced. Consistent with previous covert and side channels, the WB timing-based covert channels consist of two parties: the sender and the receiver. Similar to the attack model considered by Xiong and Szefer \cite{Xiong2020c}, we assume that the sender and receiver are co-located on the same core to share the L1 cache, and they can run in parallel on a simultaneous multithreading machine as two hyper-threads. The sender intentionally changes the cache line in a cache set to dirty status, and this can be observed by the receiver by measuring the latency of replacing the cache set. Recent research on side and covert channels \cite{Moghimi2019,Bhattacharyya2019,Aldaya2019a,Evtyushkin2018,Evtyushkin2016b,Kocher2018a,Lipp2018} has shown that it is practical for the sender and receiver to share the physical core to leak information. Moreover, there is no need for the sender and receiver to have shared memory, making the WB channel more practical. The WB time channel can be deployed not only on the L1 cache but also on other cache levels. However, this requires more operations from the sender to transfer information. In this paper, we mainly focus on the L1 cache, which is stealthier. We assume that the sender can obtain useful information about the victim, and this can be transferred to the receiver by modifying the cache line status.

\section{WB Timing-Based Channels}
In this section, we discuss how to use the dirty bits of cache lines in \textit{one} cache set to leak information, which is referred to as the \textit{target set}. Each cache line in the cache set has a dirty bit to indicate whether it has been modified so that each target set can encode multiple bits. The least-recently-used (LRU) cache replacement policy and its variants, which evict the LRU cache line, are widely deployed in Intel's modern commercial processors. For the convenience of description, we assume that the processor uses the LRU replacement algorithm. In the following sections, we will introduce the situation of real commercial processors. To transfer information using a WB channel, there are generally three phases, and these can be described as follows.

\textbf{Initialization phase.} First, the receiver performs a sequence of memory accesses to ensure that there are no dirty cache lines in the target set.

\textbf{Encoding phase.} The sender chooses whether to modify a cache line mapped to the target set to a dirty state depending on the information to be sent. Algorithms in this paper are lightweight in the encoding phase, and the sender only needs to use one cache line to encode the data.

\textbf{Decoding phase.} The receiver can conclude whether there is a dirty cache line in the target set by measuring the latency of replacing the target set.

Note that when the latency of replacing the target set is measured in the decoding phase, the target set is also initialized. Therefore, the subsequent initialization phase can be omitted. Moreover, the sender is an independent process, which means it is trivial to generate some local variables and put them in dirty states.

\begin{algorithm} [t]
	\caption{Receiver Operations in WB Channel}
	lines 0--$N$: cache lines mapping to the target set
	
	$A$, $B$: Two replacement sets without the same address
	
	$W$: Cache associativity
	
	\myline
	
	// Step 0: Initialization Phase
	
	\For {$i=0$; $i<N$; $i$++}{
		Access line $i$;
	}
	sleep; // Allow the sender code to encode \newline

	// Step 2: Decoding Phase
	
	\eIf{$times\%2$}{
		Access \textit{replacement set $A$} and accumulate time\;	
	}{
		Access \textit{replacement set $B$} and accumulate time\;
	}
	time++; // Alternate use of replacement sets
	
\end{algorithm}

Algorithms~1 and 2 show the operations of the sender and receiver in the WB channel. The sender and receiver will first agree to transfer information on a specific target set. The term \textit{lines 0--$N$} is used to denote $N+1$ different cache lines mapped to the target set. Note that \textit{line n} (where $n \in [0,N]$) represents any cache line that can be placed in the target set rather than referring to a specific physical address. There is no special constraint relationship between line~$i$ and line~$j$ ($i,j \in [0,N]$). Furthermore, the sender and the receiver are distinct Linux processes (i.e., separate programs) located in different memory spaces, which means that they share no identical cache lines.

In particular, we use the term \textit{replacement set} to refer to a set of cache lines used by the receiver to replace the target set. The replacement set typically contains $W$ cache lines, which is the same as the size of the cache set. The receiver measures the latency to access the replacement set, which will present different time distributions depending on the information transmitted by the sender. When decoding, the receiver must ensure that the cache lines in the replacement set are not in the L1 data cache, otherwise cache replacement will not happen. However, cache lines in the replacement set will remain in the L1 data cache after decoding. To figure out this limitation, the receiver can alternately operate two replacement sets with different addresses, as shown in Algorithm~2. Therefore, the cache lines used for each decoding are in the L2 cache, and after each measurement, the target set is filled with clean cache lines.

Since the L1 cache is virtually indexed and physically tagged, the receiver can easily construct a replacement set containing a collection of memory lines in its address space that all map to the target set. For an L1 cache with 64~cache sets with a cache line size of 64~bytes, bits 0--5 of the virtual address are used as the \textit{line offset}, and bits 6--11 decide the cache set. Virtual addresses with the same index bits but different tag bits will be mapped to the same cache set. 
The receiver can allocate an array of the same size as the L1 cache and select the cache lines whose index bits point to the target set, and the tag bits are different to form a replacement set.
%The receiver can allocate an array of the same size as the L1 cache but with different tag bits and select the cache lines whose index bits point to the target set to form a replacement set. 
Similarly, the sender can choose cache lines that map to the same target set. The sender and receiver can then construct covert channels according to algorithms~1 and 2.

The L1 data cache is typically an eight-way set-associative structure, which means that each cache set contains nine states of zero to eight dirty cache lines. The sender can use multiple cache lines for encoding to increase the time difference and transmission bandwidth. For binary symbol encoding, the sender can include $d_1$ or $d_2$ dirty cache lines in the target set as sending 0 or 1 ($d_1 \neq d_2$). To simplify the operations of the sender, we always choose $d_1$ equal to 0, which means that the cache line is not accessed when sending 0. For $d_2$, we can choose any value from 1 to 8 as sending 1. The more dirty cache lines the sender uses, the greater the difference the receiver can observe. However, this also requires more operations from the sender. Figure~\ref{classification}(c) shows a brief communication process between the sender and the receiver with binary-encoded symbols in which the sender transmits `10' strings. The decoding process fills the target set with clean cache lines, so no additional initialization is required. The sender can also exploit multiple cache lines in a target set to encode symbols with multiple bits. In Algorithm~1, $d$ is the number of dirty cache lines that the sender wants to include in the target set. For multi-bit symbol encoding, the sender can select a different number of dirty cache lines (a set of different values of $d$) to modulate the target set. Similarly, the receiver can extract the information by measuring the latency of accessing the replacement set.

\begin{table}[t]
	\caption{Probability of line~0 being evicted.}
	\label{table:one_probility}
	\centering
	\resizebox{0.4\textwidth}{!}
	{
		\begin{tabular}{|c|c|c|c|}
			\hline
			$N$ & LRU$^*$ & Tree-PLRU$^*$ & \begin{tabular}[c]{@{}c@{}}Intel~Xeon\\E5-2650\end{tabular} \\ \hline
			8 & 100\% & 94.3\% & 68.8\% \\ \hline
			9 & 100\% & 100\% & 81.7\% \\ \hline
			10 & 100\% & 100\% & 100\% \\ \hline
		\end{tabular}
	}
\end{table}

\subsection{Impact of the Replacement Policy} \label{section:replacement}
The LRU cache replacement policy and its variants are widely deployed in modern processors. In true LRU, the least-recently-used way is always chosen as the victim way and evicted. When the size of the replacement set is the same as that of the target set, accessing the replacement set will replace all cache lines in the target set, naturally including possible dirty cache lines. However, this is not necessarily the case for LRU variants. Consider the following memory access sequence in an eight-way data cache. Each sequence number represents a cache line mapped to the target set, and $N$ is the size of the replacement set.

\begin{itemize}
	\setlength{\itemsep}{0pt}
	\setlength{\parsep}{0pt}
	\setlength{\parskip}{0pt}
	\item Access sequence: $\overline{0}$ $\to$ $\underbrace{1 \to 2 \to ... \to N}_{Replacement\ set}$. Line~0 is a dirty cache line that maps to this cache set. Lines 1--$N$ represent $N$ different cache lines in a replacement set that also map to this cache set.
\end{itemize}

If true LRU is used, line~0 will be replaced when $N$ is equal to eight. However, the LRU algorithm needs to record the age of each cache line. For an $N$-way cache, each cache line needs $\mathrm{log}(N)$ bits to store the age, and a total of $M\times N\mathrm{log}(N)$ bits are required for a cache with $M$ cache sets. Whenever a cache line in a cache set is accessed or evicted, the LRU status of other cache lines in this cache set must be updated. Therefore, true LRU has a costly time overhead to update the LRU state and space overhead to store the age of all cache lines. According to previous reports \cite{Xiong2020c,Briongos2019}, modern high-performance processors implement approximations to LRU to save overhead, such as some pseudo-LRU policies\cite{So1988,malamy1994methods}.
% and these are known as pseudo-LRU (PLRU).

In this case, considering the above access sequence, line~0 is not guaranteed to be selected as the victim cache line in an eight-way cache when $N$ is equal to eight. However, unlike other covert channels \cite{Xiong2020c,Briongos2019} that require a profound understanding of the cache replacement policy, we noticed that no matter which replacement policy is used, accessing enough cache lines can always stably replace all cache lines in the target set.

We simulated the LRU and Tree-PLRU \cite{Zhang2011} replacement policies for an eight-way cache on gem5 \cite{Binkert2011c}, and we also conducted the same experiments on an Intel~Xeon E5-2650. In the process of these experiments, we accessed the above sequence and gradually increased the size of the replacement set; then, we recorded whether cache line~0 was replaced successfully. In each configuration, we repeated the experiment 10\,000 times, and Table~\ref{table:one_probility} shows the final results, and the processor specifications are given in Table~\ref{table:specification}.

In the PLRU replacement policy implemented on gem5, as shown in Table~\ref{table:one_probility}, the size of the replacement set needs to be at least nine to ensure that line~0 is evicted. As noted, we also tested this on an Intel~Xeon E5-2650. The experimental results show that when $N$ is at least ten, line~0 is guaranteed to be replaced.

\begin{listing}[t]
	\noindent
	\centering
	%	\begin{minipage}[t]{0.45\textwidth}
		%			\begin{lstlisting}[gobble=16, language=C,
			%	keywordstyle={\color{keywordcolor}},basicstyle = \sffamily,
			%	xleftmargin=2em,xrightmargin=2em,frame=lrtb,numberstyle={\color{numbercolor}\normalfont},	numbersep={-0.2cm},numbers=left,tabsize=7,
			%	basicstyle=\linespread{1.1}\footnotesize]
			%			rdtscp
			%			mov %eax, %esi
			%			mov (%rbx), %r8
			%			mov (%r8), %r8
			%			mov (%r8), %r8
			%			mov (%r8), %r8
			%			mov (%r8), %r8
			%			mov (%r8), %r8
			%			mov (%r8), %r8
			%			mov (%r8), %r8
			%			mov (%r8), %r8
			%			mov (%r8), %r8
			%			rstscp
			%			sub %esi, %eax
			%	\end{lstlisting}
		%	\captionof{listing}{Code used to measure replacement latency.}
		%	\end{minipage}%
	\begin{minipage}{0.48\textwidth}
		\centering
		\includegraphics[width=0.95\linewidth]{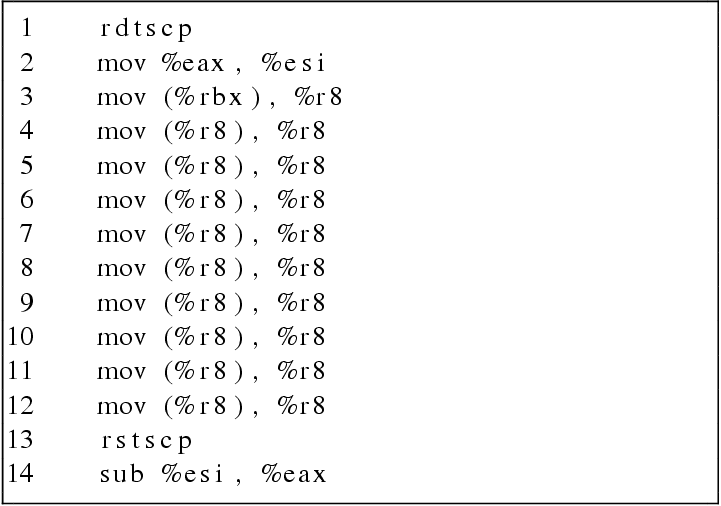}
	\end{minipage}\hfill
	\captionof{figure}{Code used to measure replacement latency.}
	\label{rep_latency}
	
	\begin{minipage}{0.48\textwidth}
		\centering
		\includegraphics[width=\linewidth]{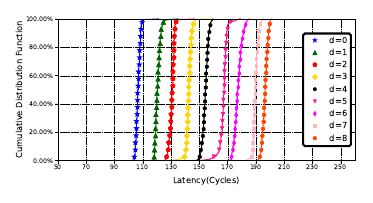}
	\end{minipage}\hfill
	\captionof{figure}{Latencies of accessing the replacement set when the target set contains different numbers of dirty cache lines.}
	\label{CDF}
\end{listing}

\subsection{Measuring the Latency} \label{section:pointer-chasing}
The receiver needs to precisely measure the latency of replacing the target set to distinguish whether there is a dirty cache line or how many dirty cache lines are in the target set. When using the \textit{rdtscp} instruction (or the \textit{lfence} and \textit{rdtsc} instructions) to measure the latency, the noise caused by serialization and the granularity of the timestamp counter will affect the accuracy of the receiver's recovery of information \cite{Osvik2006a}. The approach taken here uses the structure of the linked list and a pointer-chasing algorithm to make the measurements more accurate. We organize the replacement set into a linked list in a random order. Each element in the linked list stores the address of the next element, and the random permutation can prevent the hardware from prefetching the cache lines in the target set.

\begin{algorithm} [t]
	\caption{Covert Channel Protocol}
	$M_{s}[N]$: $N$-bit message to be transmitted by the sender
	
	$k$: the number of bits transmitted each time
	
	TSC: current time stamp counter, obtained by \textit{rdtscp}
	
	$T_s$: sender's sending period
	
	$T_r$: receiver's sampling period
	
	\myline
	
	\textbf{Sender Code:}
	
	\myline
	
	\For {$i=0$; $i<N$; $i=i+k$}{
		Encoding phase, encoding $M_{s}[i]-M_{s}[i+k-1]$;\newline
		\While{$\mathrm{TSC}<T_{\mathrm{last}} + T_s$}{
			nothing; // Allow the receiver code to decode
		}
		$T_{\mathrm{last}} = \mathrm{TSC}$;
	}

	\myline
	
	\textbf{Receiver Code:}
	
	\myline
	
	\While{True}{
		\While{$\mathrm{TSC}<T_{\mathrm{last}} + T_r$}{
			nothing; // Allow the sender code to encode
		}
		
		$T_{\mathrm{last}} = \mathrm{TSC}$;
		
		Decoding phase;
	}
\end{algorithm}

\begin{scriptsize}
	\begin{table}[t]
		\caption{Specifications of the tested CPU models.}
		\label{table:specification}
		\centering
		\begin{tabular}{|l|l|}
			\hline
			\textbf{Model} & \textbf{Intel~Xeon E5-2650}\\
			\hline
			\hline
			Microarchitecture & Sandy Bridge \\
			\hline
			Frequency & 2.2~GHz \\
			\hline
			Number of cores & 12\\
			\hline
			L1D size of each core & 32~KiB\\
			\hline
			L1D associativity & 8-way\\
			\hline
			Number of L1 cache sets & 64\\
			\hline
			Operating system & Ubuntu~16.04.12\\
			\hline
		\end{tabular}
	\end{table}
\end{scriptsize}

\begin{figure*}[t]
	\centering
	\subfloat[$d=1$]{
		\includegraphics[width=0.3\textwidth]{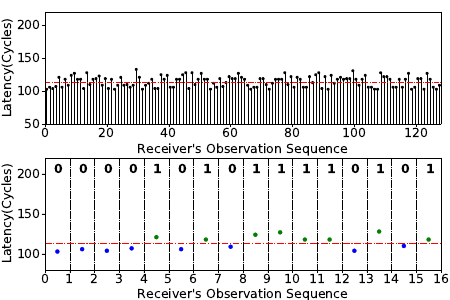}}
	\quad
	\subfloat[$d=4$]{
		\includegraphics[width=0.3\textwidth]{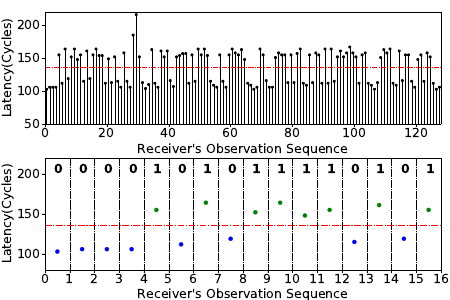}}
	\quad
	\subfloat[$d=8$]{
		\includegraphics[width=0.3\textwidth]{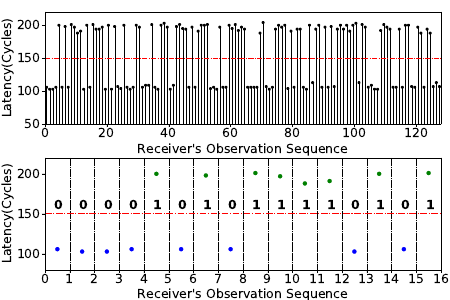}}\\
	\caption{Example sequences observed by the receiver when the sender continuously sends random sequences at a transmission rate of 400~kbps using Algorithm~3 with $T_r = 5500$ and $T_s=5500$, and (a)~$d=1$, (b)~$d=4$, and (c)~$d=8$. The upper panel in each subfigure shows a set of 128-bit random-sequence values received by the receiver, and the lower panel shows a magnified view of the reception of the first 16~bits, which are also the bits used for alignment. The dots show the latencies for the observer to access the replacement set. The dotted line in the middle represents the threshold.}
	\label{binarySeq}
\end{figure*}

\begin{figure}[t]
	\centering
	\includegraphics[width=\linewidth]{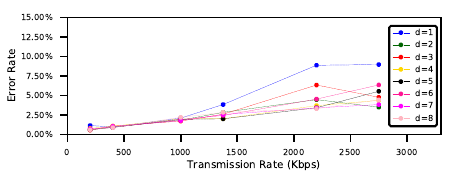}
	\caption{Relationship between transmission error rate (evaluated by edit distance) and transmission rate ($T_s=T_r$) when using Algorithm~3.}
	\label{binaryErrorRate}
\end{figure}

Figure~\ref{rep_latency} shows the assembly code used to measure the replacement latency. The \textit{rdtscp} (lines~1 and 13)~instructions can obtain the value of the time stamp counter and prevent instruction reordering and out-of-order completion. The register \textit{\%rbx} (line~3)~points to the head of the linked list. The source operand of each \textit{mov} instruction (lines~4 to 12)~depends on the data obtained from the previous \textit{mov} instruction, which makes the access to the cache line fully serialized. The measured result will finally be stored in the \textit{\%eax} register.

Figure~\ref{CDF} shows the cumulative distribution functions of the latency values of accessing the replacement set when the parameter $d=\{0,1,2,3,4,5,6,7,8\}$. For each value of $d$, we performed 1000 measurements on the Intel~Xeon E5-2650 processor, and according to Section~\ref{section:replacement}, the size of the replacement set we used was 10. As shown in Figure~\ref{CDF}, the latency values are contained in a relatively narrow band, and they are sufficiently distinguishable. This also demonstrates the feasibility of exploring the difference in the latency values of accessing the replacement set in these situations to implement time-based channels.

\section{Evaluation}
To examine the transmission rate of WB channels, we evaluated them as covert channels using one target set in the L1 data cache. As shown in Algorithm~3, the sender can encode one to multiple bits of the message $M_{s}[N]$ every $T_s$ cycles by only performing the sender operation (in Algorithm~1)~once. Then, the receiver needs to measure once using the pointer-chasing algorithm discussed in Section~\ref{section:pointer-chasing} in $T_r$ cycles. Therefore, the periods of the sender and receiver should be the same ($T_r=T_s$). 
Moreover, when using binary code symbols to transfer information, the sender can only use a single cache line to encode data, making it highly concealed. 
We evaluated WB channels presented under hyper-threaded sharing settings in the Intel~Xeon E5-2650.
In the experiments, the sender and receiver share the same physical CPU core in the form of two hyper-threads, and this can be implemented using the \textit{sched\_setaffinity} API. The sender and the receiver are distinct Linux processes (i.e., separate programs) located in different memory spaces. Additionally, we implemented covert timing channels with binary encoding symbols and multi-bit encoding symbols, which we will introduce separately.

\textbf{Symbols encoding binary.} We always assume that zero dirty cache lines in the target set ($d=0$) encodes the sending of 0. Since the L1 cache of the Intel~Xeon E5-2650 is an eight-way set-associative structure and the maximum possible number of dirty cache lines (i.e., $d$) is eight, we are not limited to using only one cache line. We analyzed situations in which parameter values $d=\{1,2,3,4,5,6,7,8\}$ were used to transmit 1. As introduced in Sections~\ref{section:replacement} and~\ref{section:pointer-chasing}, the size of the replacement set used was 10, and we measured the latency using the pointer-chasing algorithm.

Figure~\ref{binarySeq} shows example sequences observed by the receiver when the sender continuously sends random sequences using Algorithm~3. Due to limitations of space, only the results when $d=1$, $4$, and $8$ are shown in Figure~\ref{binarySeq}. When the sender is sending bit~1, the latency of accessing the replacement set by the receiver is longer due to the dirty cache lines in the target set. Furthermore, each dirty cache line increases the receiver's replacement latency by approximately 10~cycles. Therefore, the more dirty cache lines the sender contains in the target set, the more obvious the latency difference measured by the receiver, but this also requires more operations from the sender.

We also evaluated the bit error rates when $d$ takes different values. Three types of errors may occur in the transmission channel: 1)~bit flip, 2)~bit insertion, or 3)~bit loss \cite{Xiong2020c}. This paper uses the Wagner--Fischer algorithm \cite{Navarro2001} to calculate the edit distance between the sender sequence and the receiver sequence to evaluate the bit error rates. During the evaluation, the sender continuously sends a 128-bit random sequence, and the first 16~bits of the random sequence are set to a fixed value for the receiver to identify. The 128-bit random sequence is sent at least 90~times to obtain the average bit error rates. The upper panel of each subgraph in Figure~\ref{binarySeq} shows a set of 128-bit random sequence values observed by the receiver, and the lower panel shows a magnified view for the reception of the first 16~bits.

\begin{figure}[t]
	\centering
	\includegraphics[width=\linewidth]{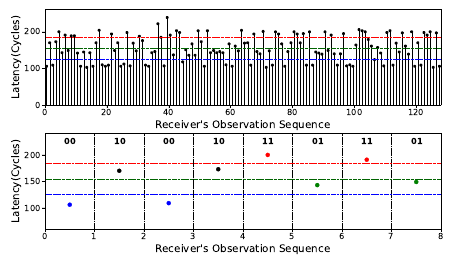}
	\caption{Example sequences observed by the receiver when the sender continuously sends random sequences at a transmission rate of 1100~kbps using Algorithm~3 with $T_r = 4000$ and $T_s=4000$. The upper panel shows a set of 256-bit random sequence values received by the receiver, and the lower panel shows a magnified view for the reception of first 16~bits, which are also the bits used for alignment. The dots show the latencies for the observer to access the replacement set. The dotted line represents the threshold.}
	\label{bitsSeq}
\end{figure}

\begin{table}[t]
	\centering
	\caption{Latency of cache access (cycles).}
	\label{table:latency}
	\begin{tabular}{|c|c|c|c|}
		\hline
		& \textbf{L1D Hit} & \textbf{\begin{tabular}[c]{@{}c@{}}L2 Hit+Replacing \\a clean cache line\end{tabular}} & \textbf{\begin{tabular}[c]{@{}c@{}}L2 Hit+Replacing \\a dirty cache line\end{tabular}} \\ \hline
		\textbf{\begin{tabular}[c]{@{}c@{}}Intel~Xeon\\E5-2650\end{tabular}} & 4--5 & 10--12 & 22--23 \\ \hline
	\end{tabular}
\end{table}

We evaluated $T_r=T_s = \{800$, $1000$, $1600$, $2200$, $5500$, $11\,000\}$, and Figure~\ref{binaryErrorRate} shows the bit error rates of the channels versus the different transmission rates. As shown in the figure, when the transmission rates are low, the values of $d$ do not affect the bit error rates much, and the increase in $d$ only allows the receiver to observe a more noticeable time difference. When the transmission rate increases to 1375~kbps (i.e., $T_s=1600$~cycles), all bit error rates are still less than 5\%. However, as the transmission rate increases further (i.e., $T_s$ decreases), we can see an observable impact on the sender--receiver communication with different $d$ values. In general, the bit error rate decreases as $d$ increases. This is because a larger $d$ makes the channel have stronger anti-interference ability and allows the receiver to distinguish the time difference more clearly. Moreover, when $d=1$, the bit error rate is significantly higher than other $d$ cases. When $d=8$, the bit error rate is only 4.5\% at a transmission rate of 2700~kbps.

In general, the error rate increases as the transmission rate increases. This is because in hyper-threaded sharing, the sender and receiver processes execute in parallel, and a higher transmission rate means a lower sending period. The sender modulates each bit of information only once and then waits for the receiver to decode it. As the sending period increases, the time during which the receiver can correctly decode is longer, and the receiver's decoding operation is more likely to occur during the sender's sleep period.

\textbf{Symbols encoding multiple bits.} The sender can set the target set to contain nine states of zero to eight dirty cache lines, which means that a theoretical maximum of three bits can be encoded. To reduce the impact of pollution caused by other processes on the target set and increase the distinction between different encoding symbols, we only encode two bits each time and avoid using adjacent $d$ to modulate the information. Figure~\ref{bitsSeq} shows the traces observed by the receiver when the sender continuously sends random sequences.

We also evaluated the bit error rates at different transmission rates when encoding two bits each time. Similarly, we evaluated $T_r = T_s = \{800$, $1000$, $1600$, $2200$, $5500$, $11\,000\}$. We chose $d=0$, $3$, $5$, and $8$ to encode $\underbracket{00}\underbracket{01}\underbracket{10}\underbracket{11}$, respectively. During the evaluation, the sender continuously sends a 256-bit (128-pair) random sequence in a loop. Similarly, the first 16~bits of the random sequence are set to a fixed value for the receiver to identify. The 256-bit random sequence is sent at least 45~times to obtain the average bit error rates. The results of our experiments show that the bit error rate is only 3.5\% at a transmission rate of 4400~kbps with symbols encoding two bits, which is significantly higher than the 1375 to 2700~kbps for encoding binary data. We note that more complex encoding mechanisms may achieve higher information transmission rates, but our goal is to illustrate a way for senders to achieve higher bandwidths.

\begin{figure*}[t]
	\centering
	\includegraphics[width=\linewidth]{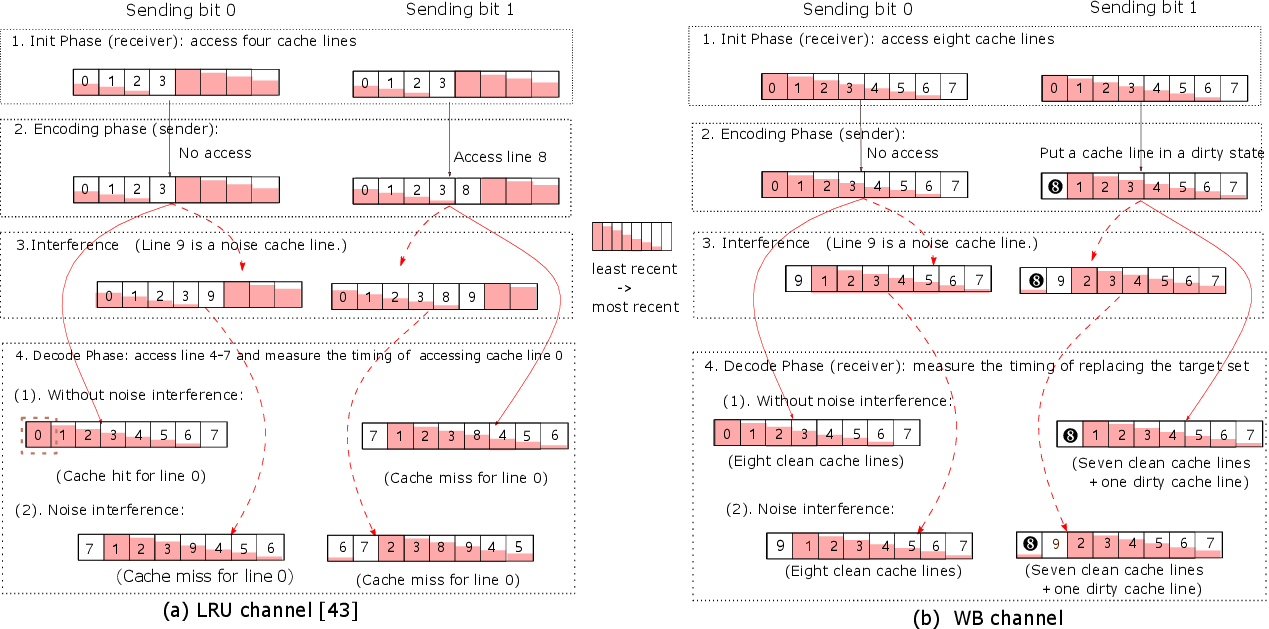}
	\caption{Effect of noisy cache lines on the LRU and WB covert channels. The numbers represent cache lines of different addresses in the target set. Shaded areas indicate the age of the cache line.}
	\label{interfence}
\end{figure*}

\section{Stability of WB Channels} \label{stability}
In most of the existing cache covert and side channels, the sender affects whether certain cache lines are available in the cache to leak information. For example, in the Flush+Reload attack \cite{Yarom2014}, the receiver first flushes a cache line from the cache. The sender fetches this cache line into the cache as transmitting bit~1, which can be observed by the receiver by measuring the time to access this cache line. Apparently, the Flush+Reload channel requires using the \textit{clflush} instruction and shared memory between the sender and receiver. However, due to security considerations, the \textit{clflush} instruction has been disabled in the Native Client sandbox \cite{Maurice2015}. Moreover, most cloud providers have disabled memory sharing across virtual machines, which naturally defends against such attacks. The Flush+Flush \cite{Gruss2016a} channel has the same requirements, but our WB channel does not.

Xiong and Szefer \cite{Xiong2020c} established a channel based on the LRU state, transmitting information without the \textit{clflush} instruction or shared memory. In the LRU channel, the sender's access to the cache line will modify the LRU state and change the eviction order of the receiver's cache line, which can be observed through a timing measurement. However, limited by the fact that any access to the target set will change the LRU state, it is difficult for the LRU channel to encode multiple bits per target set. In contrast, our WB channel has no such limitation. The peak transmission bandwidth of LRU time-based channels is 600~kbps, while our WB channels can reach 1300 to 4400~kbps, far exceeding the LRU channels. Moreover, the LRU channels are easily disturbed by the cache lines loaded into the target set by another part of the program or other processes on the core, which is referred to as a \textit{noisy cache line}, and this will lead to incorrect decoding by the receiver. 
Figure~\ref{interfence}(a) introduces the effect of a noisy cache line on the LRU channel when transmitting the information in an eight-way cache.
In the LRU channel as shown in Figure~\ref{interfence}(a), the sender changes the cache replacement state by whether to access cache line 8, which ultimately affects whether cache line 0 is in the cache.
When there is no noisy cache line, the receiver will observe an L1 cache hit when accessing line~0 if the sender is sending $m=0$, and it will observe an L1 cache miss if the sender is sending $m=1$ [steps~1, 2, and 4(1) in Figure~\ref{interfence}(a)]. However, when there is a noisy cache line(line 9 in the figure), a cache miss will always occur when accessing line~0 [steps~1, 2, 3, and 4(2) in Figure~\ref{interfence}(a)]. Therefore, it is difficult for the LRU channel to transmit information when there are other processes. Moreover, commercial processors often adopt a PLRU policy instead of a true LRU policy, which also has an impact on the LRU channel. 

Our WB channel only considers whether the cache line in the target set is in a dirty state, making it difficult for it to be disturbed. 
As shown in Figure~\ref{interfence}(b), when there is no noisy cache line, the receiver will replace eight clean cache lines in the decoding phase if the sender is sending $m=0$, and it will replace one dirty cache line and seven clean cache lines if the sender is sending $m=1$ [steps~1, 2, and 4(1) in Figure~\ref{interfence}(b)]. If there is a noisy cache line, when the sender is sending $m=0$, the noisy cache line loaded by other processes replaces the clean cache line of the target set and does not affect our measurement. When the sender is sending $m=1$, since current processors mainly use the LRU replacement algorithm or its variants, the dirty cache line loaded by the sender is the \textit{newest} in the LRU state and will be evicted at last. Therefore, the noisy cache line again does not affect our measurement [steps~1, 2, 3, and 4(2) in Figure~\ref{interfence}(b)]. 
%Other existing cache channels always focus on whether a specific memory address is in the cache, making them easy to be disturbed, but we focus on the state of the cache set.
 Moreover, our WB channel can resist the interference of multiple noisy cache lines (for example, 7 noisy cache lines are in the cache using the LRU replacement algorithm).

Prime+Probe is another covert channel that does not require shared memory. Similar to the LRU channel, it is also affected by noisy cache lines. Moreover, compared with the WB channel, in Prime+Probe \cite{Osvik2006a}, the receiver needs to access the whole set in both the prime and the probe phases. However, in the WB channel, when decoding, the target set is also initialized at the same time, and the receiver does not need to access the whole set again.

Table~\ref{table:latency} shows the latency of cache access measured in an Intel~Xeon E5-2650. An L1 hit takes less than five CPU cycles; an L2 hit and replacing a clean cache line in the L1 data cache takes less than 12 CPU cycles. However, an L2 hit and replacing a dirty cache line in the L1 data cache takes about 22 CPU cycles. Most of the previous cache channels exploit the timing difference between cache hits and cache misses. However, our experimental results show that replacing cache lines in different states results in a more significant latency difference: almost twice the latency difference between an L1 hit and an L1 miss. Strong resistance to interference and a more distinguishable latency difference allow our WB channels to reach a transmission rate of 1300 to 4400~kbps, far exceeding other timing channels in caches \cite{Liu2015,Xiong2020c,Yao2018}. Of course, if other processes modify a cache line mapped to the target set, this will affect our WB channel. However, compared to direct access to the cache line, this is not common, and other known cache covert channel attacks also have this noise source \cite{Liu2015,Xiong2020c,Yao2018,Osvik2006a}.

\subsection{Random replacement algorithm}\label{chapter:random}
Time-based channels related to replacement policies are often based on the sender and receiver sharing the replacement state to transmit information \cite{Xiong2020c,Briongos2019}, which usually requires a profound understanding of the replacement algorithm used. However, the random replacement policy does not need any state information. When using the random replacement policy, a random cache line is selected as the victim cache line and evicted, which naturally defeats these attacks. Compared with the LRU policy, the random replacement policy gives slight degradation in cache miss rate and cycles per instruction in the SPEC~2006 int and float benchmarks \cite{Xiong2020c}. However, the random replacement policy also has the advantages of low cost and simple implementation. Most ARM processors use a pseudo-random replacement policy.

In the WB channel, the sender modulates single-bit information to whether the target set contains dirty cache lines. The impact of different replacement policies is whether the dirty cache line can be selected as the victim cache line when the receiver accesses the replacement set. Since we use the entire replacement set for replacement, even if the L1 data cache adopts a random replacement policy, the dirty cache lines still have a high probability of being replaced, $p = 1-(\frac{W-d}{W})^{L}$.

The above formula shows the probability that at least a dirty cache line in the target set can be replaced by the replacement set when the data cache adopts a random replacement policy. Here, \textit{W} represents the associativity of the data cache, $d$ represents the number of dirty cache lines in the target set, and $L$ represents the size of the replacement set. The probability increases as $d$ or $L$ increase. According to this formula, the probability is approximately equal to 99.1\% when $d=3$ and $L=10$. Therefore, simply adopting a random replacement policy still cannot effectively defeat the WB channel.

We simulated a pseudo-random replacement policy for an eight-way cache on gem5. We constructed an experiment similar to that in Section~\ref{section:replacement}. The difference was that we used more dirty cache lines and a larger replacement set. In the sequence, we accessed $d$ dirty cache lines in a loop to ensure they were in the target set. Then, we accessed the replacement set, and the probability of at least one dirty cache line being replaced was measured and recorded. We repeated the test 10\,000 times under distinct replacement set sizes and configurations with different numbers of dirty cache lines, and the final results are shown in Table~\ref{table:multi_probability}. The results show that using appropriate $d$ and $L$ values (for example, $d=3$ and $L=12$), a stable WB channel can be established in a cache using a random replacement policy. 
%Therefore, the WB channel does not require a profound understanding of the replacement algorithm.

\begin{scriptsize}
	\begin{table}[t]
		\centering
		\caption{Probability of at least one dirty cache line being replaced.}
		\label{table:multi_probability}
		\resizebox{0.46\textwidth}{!}{
			\begin{tabular}{|c|c|c|c|c|c|c|}
				\hline
				& $L=8$ & $L=9$ & $L=10$ & $L=11$ & $L=12$ & $L=13$ \\ \hline
				$d=2$ & 63.6\% & 75.9\% & 84.6\% & 89.0\% & 92.9\% & 95.0\% \\ \hline
				$d=3$ & 89.5\% & 94.4\% & 96.8\% & 98.3\% & 99.4\% & 99.5\% \\ \hline
			\end{tabular}
		}
	\end{table}
\end{scriptsize}

In the Prime+Probe attack, when the processor uses a random replacement policy, it is difficult for the receiver to completely fill the target set with known cache lines in the prime phase. Moreover, on processors using the LRU policy, using the reverse traversal order in the prime and probe phases may avoid thrashing, i.e., self-eviction by the receiver's own data \cite{Liu2015}. However, this does not work on a cache using a random replacement policy, and 0 to 8 cache misses may be measured in the probe phase.

\section{Stealthiness of WB Channels}
The LRU channel requires the sender to constantly modulate the transmitted bit (accessing the cache line) within the encoding time of $T_s$, and the receiver performs a decoding operation every $T_r$ time ($T_r < T_s$). However, uneven thread scheduling leads to the fact that for each bit of information, the receiver does not always perform $T_s/T_r$ measurements, which increases both the difficulty of decoding and the bit error rate. However, each bit of information in our WB channel only needs to be modulated once, making it more straightforward. Table~\ref{table:load_num} shows the number of cache loads per millisecond measured from the hardware performance counter using the Linux perf tool when $T_s=11\,000$. As shown in Table~\ref{table:load_num}, the total number of cache loads per millisecond of our WB channel is about 59.8\% of that of the LRU channel, making our channel stealthier.

\begin{table*}[h]
	\begin{minipage}[t]{0.3\linewidth}
		\centering
		\caption{Cache load numbers per millisecond of the sender process.}
		\label{table:load_num}
		
		\begin{tabular}{|c|c|c|}
			\hline
			& WB & LRU \\ \hline
			L1 & $3.151\times 10^8$ & $5.265\times 10^8$ \\ \hline
			L2 & $1.217\times 10^5$ & $6.840\times 10^4$ \\ \hline
			LLC & $2.203\times 10^3$ & $2.213\times 10^3$ \\ \hline
			Total & $3.153\times 10^8$ & $5.266\times 10^8$ \\ \hline
		\end{tabular}
	\end{minipage}
	\begin{minipage}[t]{0.7\linewidth}
		\caption{Cache miss rate of the sender process in Algorithm~1.}
		\label{table:miss_rate}
		\resizebox{\textwidth}{!}{ 	
			\centering
			\begin{tabular}{|c|c|c|c|c|c|c|c|}
				\hline
				& & \multicolumn{3}{c|}{\textbf{Binary encoding}} & \multicolumn{3}{c|}{\textbf{Multi-bit encoding}} \\ \hline
				& & \textbf{L1 WB} & \textbf{\begin{tabular}[c]{@{}c@{}}Sender\\ \& g++\end{tabular}} & \textbf{\begin{tabular}[c]{@{}c@{}}Sender\\ only\end{tabular}} & \textbf{L1 WB} & \textbf{\begin{tabular}[c]{@{}c@{}}Sender\\ \& g++\end{tabular}} & \textbf{\begin{tabular}[c]{@{}c@{}}Sender\\ only\end{tabular}} \\ \hline
				\multirow{3}{*}{\textbf{\begin{tabular}[c]{@{}c@{}}Intel~Xeon\\ E5-2650\end{tabular}}} & \textbf{L1D} & 0.04\% & 0.16\% & 0.003\% & 0.30\% & 0.34\% & 0.003\% \\ \cline{2-8}
				& \textbf{L2} & 3.59\% & 26.84\% & 35.16\% & 0.42\% & 15.15\% & 26.46\% \\ \cline{2-8}
				& \textbf{LLC} & 34.38\% & 2.23\% & 34.42\% & 39.08\% & 1.96\% & 35.29\% \\ \hline
			\end{tabular}
			
		}
	\end{minipage}%
	
\end{table*}

%\begin{scriptsize}
%\begin{table*}[htb]
%	\caption{Cache load numbers per millisecond of the sender process.}
%	\label{table:load_num}
%	\centering
%	\begin{tabular}{|c|c|c|}
	%		\hline
	%		& WB & LRU \\ \hline
	%		L1 & $3.151\times 10^8$ & $5.265\times 10^8$ \\ \hline
	%		L2 & $1.217\times 10^5$ & $6.840\times 10^4$ \\ \hline
	%		LLC & $2.203\times 10^3$ & $2.213\times 10^3$ \\ \hline
	%		Total & $3.153\times 10^8$ & $5.266\times 10^8$ \\ \hline
	%	\end{tabular}
%\end{table*}
%\end{scriptsize}
%
%\begin{table*}[htb]
%	\caption{Cache miss rate of the sender process in Algorithm~1.}
%	\label{table:miss_rate}
%	\centering
%		\begin{tabular}{|c|c|c|c|c|c|c|c|}
	%			\hline
	%			& & \multicolumn{3}{c|}{\textbf{Binary encoding}} & \multicolumn{3}{c|}{\textbf{Multi-bit encoding}} \\ \hline
	%			& & \textbf{L1 WB} & \textbf{\begin{tabular}[c]{@{}c@{}}Sender\\ \& g++\end{tabular}} & \textbf{\begin{tabular}[c]{@{}c@{}}Sender\\ only\end{tabular}} & \textbf{L1 WB} & \textbf{\begin{tabular}[c]{@{}c@{}}Sender\\ \& g++\end{tabular}} & \textbf{\begin{tabular}[c]{@{}c@{}}Sender\\ only\end{tabular}} \\ \hline
	%		\multirow{3}{*}{\textbf{\begin{tabular}[c]{@{}c@{}}Intel~Xeon\\ E5-2650\end{tabular}}} & \textbf{L1D} & 0.04\% & 0.16\% & 0.003\% & 0.30\% & 0.34\% & 0.003\% \\ \cline{2-8}
	%		& \textbf{L2} & 3.59\% & 26.84\% & 35.16\% & 0.42\% & 15.15\% & 26.46\% \\ \cline{2-8}
	%		& \textbf{LLC} & 34.38\% & 2.23\% & 34.42\% & 39.08\% & 1.96\% & 35.29\% \\ \hline
	%	\end{tabular}
%\end{table*}

Table~\ref{table:miss_rate} shows the cache miss rates of the sender process in Algorithm~1 measured from the hardware performance counters using the Linux perf tool. To provide a baseline with no attack, we measured the result of only the sender process on the physical core (denoted by \textit{sender only}) and the result of the sender process sharing the physical core in the form of hyper-threading with a benign \textit{g++} workload (denoted by \textit{sender \& g++}), similar to the work of Xiong et~al. \cite{Xiong2020c}.

When there is only the sender process, the L1 cache miss rate is the smallest. However, due to reduced access to L2 and LLC, it still has a relatively high miss rate for L2 and LLC. In the WB channel, the receiver process needs to measure the latency of accessing the replacement set, which causes the cache lines of the sender process in the target set to frequently be evicted to the L2 cache. Therefore, the sender process in WB channels has a higher L1 cache miss rate and a lower L2 cache miss rate. The WB channel also has a relatively high LLC miss rate due to fewer references to the LLC.

Compared with a sender using one dirty cache line for binary encoding, a sender using multi-bit encoding symbols needs to modulate two bits of information to a different number of dirty cache lines, making it have a higher L1 cache miss rate. However, although the WB channel uses the cache line replacement timing difference to leak information, the sender process only modulates each bit of information once, and the receiver only replaces the cache lines in the target set. The sender and receiver processes spend most of their time in sleep waiting for encoding or decoding. Therefore, when a benign program such as \emph{g++} shares a physical core with the sender process, it will cause even greater cache contention than that due to the receiver in the WB channel. Hence, if a victim wants to use performance counters to detect possible time-based channels \cite{Chiappetta2016,Zhang2016a,Alam2017}, the WB channel is difficult to distinguish from contention due to benign programs.

\section{Defending Against WB Channels}
Compared with software solutions to mitigate cache channel attacks, secure caches have a minor performance loss. However, most previous secure cache target attacks exploit the timing difference between cache hits and cache misses. We introduce a new classification of cache channels to emphasize that the design of secure caches should be more comprehensive. The current main security cache methods are noise injection \cite{Hu1992,Glemser1955,Fang2018,Fang2021}, randomization \cite{Evtyushkin2016a,Ssociativity1997,Wang2007a,Liu2015a}, and partitioning \cite{Kiriansky2018b,Domnitser2012a,Lee2005,Strackx2010,Wang2007a}.

\textbf{Noise injection.} Fang et~al. proposed Prefetch-guard \cite{Fang2018,Fang2021}, which leverages a hardware prefetcher to inject noise into the cache sets involved in the attack, and this is effective against Flush+Reload, Prime+Probe, and other attacks. However, as described in Section~\ref{stability}, the noisy cache lines prefetched by Prefetch-guard cannot effectively defend against the WB channel. Introducing noise to reduce the resolution of the clock \cite{Hu1992,Glemser1955} can also be used as a countermeasure against the attack. However, this method may affect benign applications that require high-precision clocks. Moreover, an attacker can use other methods to generate high-resolution clocks. For example, running a clock process in a separate execution core \cite{Yarom2014}.

\textbf{Randomization.} 
The random fill cache \cite{Liu2015a} de-correlates the cache fill and the demand for memory access. When a cache miss occurs, the missed data is directly sent to the processor without filling the cache. The cache is filled with random fetches in a configurable neighbor window of the missing memory line instead. 
Although if the cache line is already in the cache, modifying it can still change the cache line to a dirty state. 
The receiver’s access does not guarantee that the desired cache line to be fetched into cache, and the cache line fetched in the random window may not necessarily be mapped to the target set. Therefore, when the random window is large enough, the RF cache can effectively mitigate the WB channel.
As noted in Section~\ref{chapter:random}, simply adopting a random replacement policy cannot effectively defend against the WB channel. Other cache randomization techniques \cite{Qureshi2018a,Ssociativity1997,Wang2007a,ZSong2021} remove any discernible relation between co-resident lines in a set by randomizing the mapping relationship between addresses and cache sets. However, fixed random mapping can still leak information. Although frequently updating the mapping relationship can increase security, the updating frequency requires further consideration of the trade-off between performance and security.

%However, if the cache line is already in the cache, modifying it can still change the cache line into a dirty state. 
%The cache line accessed by the receiver cannot be guaranteed to be fetched in the cache, and the cache line fetched in the random window may not necessarily be mapped to the target set.
%
%the receiver’s multiple rounds of access does not guarantee that the desired cache line to be fetched into cache

%Moreover, the WB channel does not target specific memory addresses, and random fetching of cache lines will still cause cache replacement, which can be measured equivalently by the receiver. 
%
%Therefore, a random fill cache can effectively mitigate Flush+Reload, Prime+Probe, and the LRU channel without shared memory, but it is invalid for our WB channel. 

\textbf{Partitioning the cache.} In this defense, the cache is statically or dynamically partitioned into different regions for different processes, and each process can only access the cache lines in its region. Static partitioning, e.g., NoMo \cite{Domnitser2012a}, reserves several ways of a set for each hardware process, and this always results in a significant performance loss. Dynamic partitioning, e.g., PLcache \cite{Wang2007a}, locks protected cache lines in the cache set and does not allow them to be evicted by other processes. PLcache is effective for mitigating the WB channel, because when the dirty cache line is locked, the receiver cannot replace it during the decoding phase. However, PLcache may cause excessive locking because it does not support the replacement of locked cache lines, even when they are switched out and not active \cite{Kong2009}. Other approaches, such as DAWG \cite{Kiriansky2018b}, also propose to partition the cache to eviction isolation, which also mitigates WB channels.

Using write-through caches can effectively defend against the WB channel because data is updated synchronously to the cache and main memory, 
and the cache line does not need a dirty bit to record whether it has been modified. 
However, this is not a general method to defend against cache attacks. Moreover, compared to the write-back cache, a write-through cache results in significant CPU performance degradation \cite{Dai2011}, making it not been widely adopted in commercial processors.

\begin{figure}[t]
	\centering
	\includegraphics[width=0.95\linewidth]{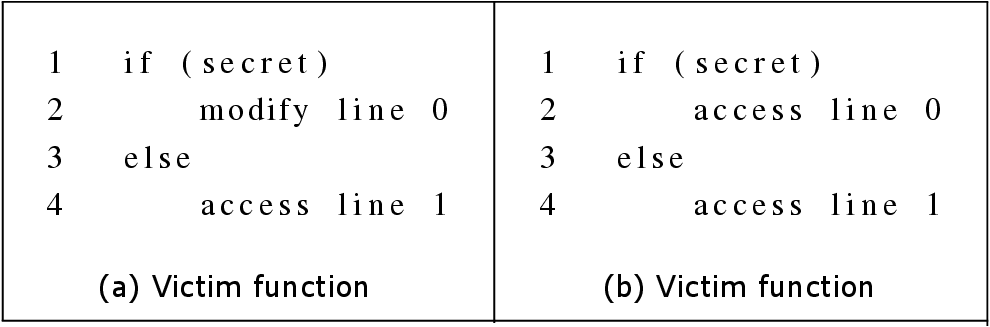}
	\caption{WB gadgets for side channel exploitation}
	\label{Victim function}
\end{figure}

\section{Side-Channel Attack}
This section describes how our WB channel can be extended to possible side channels when the victim has secret-dependent access. Figure~\ref{Victim function} shows possible gadgets where line~0 is in the cache set $m$ and line~1 is in the cache set $n$. There are usually three scenarios for information leakage through side channels. 1)~As shown in Figure~\ref{Victim function}(a), the data is modified in the branch, which changes the cache line into a dirty state. By measuring the latency of replacing set $m$, we can distinguish whether the \textit{secret} is 1 or 0. In this scenario, it does not matter whether line~0 and line~1 are the same cache line, different cache lines in the same cache set, or different cache lines in different cache sets. However, the Prime+Probe and LRU channels do not work when line~0 and line~1 are in the same cache set. 2)~In many encryption algorithms, the private key is read-only and cannot be modified. When the data is not modified in the branch, we can use a similar approach to Prime+Probe. As shown in Figure~\ref{Victim function}(b), the attacker fills set $m$ with $W$ dirty cache lines. If the secret is 1, a dirty cache line in set $m$ will be replaced. Therefore, the secret can also be inferred by measuring the latency of replacing set $m$. In this case, line~0 and line~1 need to belong to different cache sets. 3)~The attacker can also affect the execution time of the victim function to leak information. As shown in Figure~\ref{Victim function}(b), the attacker first fills set $m$ with $W$ dirty cache lines and set $n$ with $W$ clean cache lines. If the secret is 1, when cache line~0 is accessed, a dirty cache line in set $m$ needs to be written back to the backing store, which takes a longer time. The attacker can infer the secret by measuring the latency to call the victim's function. In the first two scenarios, the execution of the victim function affects the state of the cache set. The attacker can use the pointer-chasing algorithm (Section~\ref{section:pointer-chasing}) to reduce noise interference and effectively extract the secret. However, in the third scenario, the time difference in calling the victim function can easily be overwhelmed by noise. The experimental results show that only when each branch loads two cache lines serially in the cache set can the attacker clearly observe the time difference and infer the secret.

%clearly observe the time difference and infer the secret.

%\captionsetup[subfigure]{labelformat=parens,labelsep = none}
%\begin{listing}[t]
%	\noindent
%	\centering
%	\begin{minipage}{0.24\textwidth}
	%		\begin{framed}
		%			\begin{lstlisting}[gobble=16, language=C,numbers=left,
			%			numberstyle={\color{numbercolor}\normalfont},
			%			numbersep={-0.2cm},keywordstyle={\color{keywordcolor}},
			%			breaklines=true,
			%			basicstyle=\linespread{1.3}\footnotesize,
			%			aboveskip=0em,
			%			belowskip=0em,tabsize=6
			%			]
			%			if (secret)
			%			modify line 0
			%			else
			%			access line 1
			%			\end{lstlisting}
		%%			\captionof{subfigure}{ Victim function}
		%		\end{framed}
	%	\end{minipage}%
%	\begin{minipage}{0.24\textwidth}
	%		\begin{framed}
		%			
		%			\begin{lstlisting}[gobble=16, language=C,numbers=left,
			%			numberstyle={\color{numbercolor}\normalfont},
			%			numbersep={-0.2cm},keywordstyle={\color{keywordcolor}},
			%			breaklines=true,
			%			basicstyle=\linespread{1.3}\footnotesize,
			%			aboveskip=0em,
			%			belowskip=0em,tabsize=6
			%			]
			%			if (secret)
			%			access line 0
			%			else
			%			access line 1
			%			\end{lstlisting}
		%%			\captionof{subfigure}{ Victim function}
		%		\end{framed}
	%	\end{minipage}%
%	\caption{WB gadgets for side channel exploitation}
%\end{listing}

\section{Conclusions}
In this paper, we have highlighted that using cache misses alone may result in more significant time differences than those between cache hits and cache misses, and we propose a new and broader classification of cache covert channel attacks. Moreover, we have presented novel cache channels leveraging the cache line dirty states on a real-world commercial processor. We implemented timing channels with binary encoding symbols and multiple-bit encoding symbols without the need for shared memory between the sender and the receiver. The experimental results show that the peak transmission bandwidth of the WB channel can vary from 1300 to 4400~kbps, far exceeding other time-based cache channels. Unlike most existing cache channels, which always target specific memory addresses, the new WB channels focus on the cache set and cache line states, making the channel more resistant to interference. Moreover, the WB channel has fewer cache load operations and can still work in a cache using a random replacement policy. In addition, it is difficult to effectively detect the WB channel by monitoring the cache miss rate. We have proposed several methods to mitigate the WB channel, and we have also introduced ways in which to use it for side-channel attacks.

\section*{Acknowledgment}

We would like to thank the authors of the LRU channel \cite{Xiong2020c}, especially Wenjie Xiong. Thanks for her willingness to open her source code and the help provided.

\bibliographystyle{IEEEtranS}
\bibliography{IEEEabrv,refs}

%\begin{thebibliography}{1}
%\end{thebibliography}

\end{document}